\PassOptionsToPackage{dvipsnames,prologue}{xcolor}

\documentclass[acmsmall,nonacm]{acmart}

\acmJournal{PACMPL}
\acmVolume{1}
\acmNumber{X} 
\acmArticle{1}
\acmYear{2024}
\acmMonth{5}
\acmDOI{} 
\startPage{1}

\setcopyright{none}

\usepackage{booktabs}   
\usepackage{subcaption} 

\usepackage{listings}
\usepackage[dvipsnames]{xcolor}
\usepackage{mdframed} 
\usepackage{xcolor}
\usepackage{calc}
\usepackage{enumitem}

\usepackage{amsmath}

\newcommand{\N}{\mathbb{N}}
\newcommand{\E}{\exists}

\definecolor{light-gray}{gray}{0.95}
\definecolor{deep-gray}{gray}{0.80}

\newcommand{\hc}[1]{\colorbox{deep-gray}{#1}}
\newcommand{\slf}{\hspace{-2pt}}

\newcommand{\zero}{\texttt{Zero}}

\begin{document}

\title{Don't Call Us, We'll Call You}         
\subtitle{Towards Mixed-Initiative Interactive Proof Assistants for Programming Language Theory}                     


\author{Jan Liam Verter}
\affiliation{
  \position{Position1}
  \department{Department of Distributed and Dependable Systems}              
  \institution{Faculty of Mathematics and Physics, Charles University}            
  \country{Czech Republic}                    
}
\email{verter@d3s.mff.cuni.cz}          

\author{Tomas Petricek}
\affiliation{
  \department{Department of Distributed and Dependable Systems}              
  \institution{Faculty of Mathematics and Physics, Charles University}            
  \country{Czech Republic}                    
}
\email{tomas@tomasp.net}          


\begin{abstract}
There are two kinds of systems that programming language researchers use for their work. Semantics
engineering tools let them interactively explore their definitions, while proof assistants can
be used to check the proofs of their properties. The disconnect between the two kinds of systems
leads to errors in accepted publications and also limits the modes of interaction available when
writing proofs.

When constructing a proof, one typically states the property and then develops the proof manually
until an automatic strategy can fill the remaining gaps. We believe that an integrated and more
interactive tool that leverages the typical structure of programming language could do better.
A proof assistant aware of the typical structure of programming language proofs could require less
human input, assist the user in understanding their proofs, but also use insights from the
exploration of executable semantics in proof construction.

In the early work presented in this paper, we focus on the problem of interacting with a proof
assistant and leave the semantics engineering part to the future. Rather than starting with manual
proof construction and then completing the last steps automatically, we propose a way of working
where the tool starts with an automatic proof search and then breaks when it requires feedback
from the user. We build a small proof assistant that follows this mode of interaction and
illustrates the idea using a simple proof of the commutativity of the ``+'' operation for Peano
arithmetic. Our early experience suggests that this way of working can make proof construction easier.
\end{abstract}



\maketitle


\section{Introduction}

Programming language researchers use two kinds of tools to describe the semantics of
programming languages and study their properties. Semantics engineering tools like PLT Redex
\cite{felleisen-2009-semantics} make it possible to define, explore and debug the operational
semantics of a language, while proof assistants such as SASyLF \cite{aldrich-2008-sasylf},
Coq and Agda are used to formalize and check proofs of programming language properties.
Unfortunately, there is a disconnect between these two kinds of tools, leading to errors even
in peer-reviewed high-profile publications \cite{klein-2012-run}. Our research objective is to
design a unified system that will integrate support for programming language semantics description, inspired by PLT Redex, with support for proof checking, inspired by SASyLF. We are interested
in exploring research questions such as:
\begin{enumerate}
\item Can the proof assistant leverage the typical structure of programming language proofs to
  support a more effective interactive proof development?
\item Can we use interactive exploration of executable semantics as the basis for constructing
  proofs, e.g., by taking inspiration from programming by demonstration \cite{cypher-1993-watch}?
\item Can we make writing programming language models and proofs more accessible by
  writing them directly in the concrete language syntax \cite{storm-2023-concrete}?
\end{enumerate}
In this paper, we focus on the first question and present our early work on developing a proof
assistant that is based on a novel mode of interaction where the assistant initiates
the search for a proof using a very simple proof strategy, but asks the user for guidance
when a more sophisticated human insight is needed. We draw on human-computer interaction research,
where the approach where ``each agent (human or computer) contributes what [it is best equipped
for] at the most appropriate time'' is known as mixed-initiative interaction \cite{allen-1999-mixed}.

Ideally, the resulting interaction should mimick the level of detail needed in
rigorous informal proofs in programming language theory. Those typically consist of instructions
such as what kind of induction to use (e.g., over term structure, over typing derivation),
what induction hypothesis to use, how to use a lemma, how to do a case splitting and
further hints for non-trivial cases.

To illustrate the idea, we developed a proof assistant for first-order predicate logic based
on the natural deduction system. The assistant can check proofs written in the explicit
Fitch-style notation, but it can also semi-automatically complete partial proofs by
interacting with the user. In this paper, we demonstrate the process using a simple
proof of commutativity of the ``+'' operation for Peano arithmetic. The proof itself is
of little interest, but it shows the integration between automatic search and human input
that we aim to support for typical programming language research proofs.

\section{Worked example}

The mode of interaction with a typical proof assistant is that the user manually constructs
the proof structure and, after providing enough structure, invokes an ``auto'' command to
complete the proof through search. In this section, we introduce an alternative mixed-initiative
mode of interaction where the user invokes the proof assistant, which proceeds with automatic
search, but asks the user when it needs more input to complete the proof.
To contrast the two modes of interaction, we use a simple theorem in the theory of natural numbers.
This worked example, although simplistic, shows how the intended interaction proceeds.

We first give a description of the proof in prose to show how such a theorem might be justified
in a research paper (Section~\ref{sec:informal}). This demonstrates the proof structure and
the level of detail that both, the writer and the reader of the justification might consider
satisfactory. We then try to match that level of detail in the formal proof presented (Section~\ref{sec:formal})
and show how a proof with the same level of detail can be constructed using our proposed mixed-initiative
mode of interaction (Section~\ref{sec:interactive}).

\subsection{A Primer on Notation}
First, we showcase the notation of our tool using a simple example. The following derives the modus ponens theorem schema using the built-in implication elimination:

\vspace{0.5em}
\begin{mdframed}[backgroundcolor=light-gray, roundcorner=10pt,leftmargin=0, rightmargin=0, innerleftmargin=0, innertopmargin=0,innerbottommargin=0, outerlinewidth=0, linecolor=light-gray]
\begin{minipage}{\linewidth}
\begin{lstlisting}[mathescape=true,columns=flexible,breaklines=true,breakatwhitespace=true,escapechar=^]
theorem schema modus-ponens for all propositions P, K  : P $\implies$ K, P $\vdash$ K
proof : | p$\rightarrow$k : P $\implies$ K
        | p : P
        ^\hspace{1.2pt}^|------------------------------
        | K  by rule $\implies$^\hspace{-3pt}^-elim on p$\rightarrow$k, p
\end{lstlisting}
\end{minipage}
\end{mdframed}
\vspace{0.5em}

The example demonstrates the notation well -- we can label assertions and sub-proofs with names (such as \texttt{p$\rightarrow$k} and \texttt{proof}), those are then used for referring to them later. Note that \texttt{p$\rightarrow$k} is just a conveniently named identifier and we could equally write, e.g. \texttt{p\_imp\_k}. We can also omit the naming of the statement when we don't need to directly address it. For example, we omit naming the last statement in the sub-proof, because it is implicitly used as the result.

The tool has built-in support for all logical connectives and in the context of defining a theorem (schema) it also supports the ``turnstile''. This convenient notation allows us to use a theorem defined like this one (with one or more premises and a turnstile) using only a single justification without having to first claim the statement it represents and then use one or more implication eliminations to obtain the conclusion. However, nothing prevents us from using the more explicit notation and defining it explicitly as an implication: \texttt{(P$\implies$K)$\implies$P$\implies$K} -- both declarations are proved the same way.

We use built-in rules for the introduction and elimination of the usual logical connectives to justify logical statements. To make the above example more interesting, we also employ the limited support for parametric theorems and rules (parametrized over propositions) that is available in our prototype. However, we will not make use of this facility in the main example.

\subsection{Informal Proof}
\label{sec:informal}

The example that we use to explain the proposed mode of interaction is the following formula:

\vspace{0.5em}
\texttt{$\forall$ (n$_1$ : $\N$) (n$_2$ : $\N$) : $\E$ (n$_3$ : $\N$) : Sum(n$_1$, n$_2$, n$_3$) $\land$ Sum(n$_2$, n$_1$, n$_3$)}
\vspace{0.5em}

\noindent
The justification in the prose form might typically look like this:

\noindent\textit{The proof proceeds by induction on the first variable (\texttt{n}$_1$).}
\begin{enumerate}[leftmargin=15pt]
  \item \textit{When \texttt{n$_1$} is \texttt{Zero}, the corresponding goal \texttt{Sum(Zero, n$_2$, N$_3$) $\land$ Sum(n$_2$, Zero, N$_3$)} for an unspecified (universal) \texttt{n$_2 : \N$} and some arbitrary \texttt{N$_3 : \N$} is justified using \texttt{sum-zero} for the left-hand side and an auxiliary lemma \texttt{$\forall$ (n : $\N$) : Sum(n, Zero, n)} for the right-hand side.}

  \item \textit{When \texttt{n$_1$} is \texttt{S(m$_1$)} for an unspecified (universal) \texttt{m$_1$}, we obtain an induction hypothesis\\\texttt{$\forall$ (n$_2 : \N$) : $\E$ (n$_3 : \N$) : Sum(m$_1$, n$_2$, n$_3$) $\land$ Sum(n$_2$, m$_1$, n$_3$)}. We then use case analysis on the universal variable \texttt{n$_2$} splitting the proof into two cases:}

  \begin{enumerate}[leftmargin=15pt]
    \item \textit{When the \texttt{n$_2$} is \zero, we have the goal \texttt{Sum(n$_1$, Zero, N$_3$) $\land$ Sum(Zero, n$_1$, N$_3$)} for some arbitrary \texttt{N$_3 : \N$}. We solve it by following the same steps as above only switching the order -- we use the auxiliary lemma to justify the left-hand side and the rule \texttt{sum-zero} to justify the right-hand side of the conjunction.}

    \item \textit{When the \texttt{n$_2$} is \texttt{S(m$_2$)} for an unspecified (universal) \texttt{m$_2$ : $\N$}, we finally get to the interesting part. To justify our goal \texttt{Sum(S(m$_1$), S(m$_2$), N$_3$) $\land$ Sum(S(m$_2$), S(m$_1$), N$_3$)} for some arbitrary \texttt{N$_3 : \N$} we use the induction hypothesis on \texttt{S(m$_2$)} obtaining \texttt{$\E$ (n$_3 : \N$) : Sum(m$_1$, S(m$_2$), n$_3$) $\land$ Sum(S(m$_2$), m$_1$, n$_3$)}. Next we use the rule \texttt{sum-s} on the left-hand side and the auxiliary lemma \texttt{$\forall$ (n$_1$,n$_2$,n$_3 : \N$) : Sum(n$_1$, n$_2$, n$_3$) $\implies$ Sum(n$_1$, S(n$_2$), S(n$_3$))} on the right-hand side of its "instantiation". This gives the justification for our final goal.}

  \end{enumerate}
\end{enumerate}

\subsection{Formal Proof in Checking Mode}
\label{sec:formal}
Ideally, the proof written using a proof assistant would have the same structure as the hand-written informal proof and the proof assistant would be able to complete the easy cases automatically, only requiring guidance in the trickier parts of the proof, e.g., the (2b) case.

If we follow the standard mode of interaction with a proof assistant, we write the structure of the proof ourselves and let the assistant handle the repetitive and uninteresting parts of the proof automatically using the \texttt{prove} command. When written in this way, the formal proof has a similar level of detail as the informal proof -- it keeps the focus on the parts that are interesting and require creativity -- but elaborating the structure of the proof is somewhat tedious. (This would be even more the case for programming language theory proofs that often have a typical fixed structure.)

The formal version of the proof checkable by our prototype is shown in figure \ref{fig:formal}.\footnote{The source file containing the complete module -- the proof together with definitions can be found at \href{https://gist.github.com/lambduli/a11407156b8c9c0d54d9ea6972efc73d}{this GitHub Gist}.} (With the most interesting parts requiring human creativity highlighted.)
\lstset{basicstyle=\footnotesize\ttfamily}
\begin{figure}
\begin{mdframed}[backgroundcolor=light-gray, roundcorner=10pt,leftmargin=0, rightmargin=0, innerleftmargin=0, innertopmargin=0,innerbottommargin=0, outerlinewidth=0, linecolor=light-gray]
\begin{lstlisting}[mathescape=true,columns=flexible,breaklines=true,breakatwhitespace=true,escapechar=^]
theorem sum-total-comm : $\forall$ (n$_1$, n$_2$ : $\N$) : $\exists$ (n$_3$ : $\N$) : Sum(n$_1$, n$_2$, n$_3$) $\land$ Sum(n$_2$, n$_1$, n$_3$)
$\forall$ (n$_1$ : $\N$) (n$_2$ : $\N$) : $\exists$ (n$_3$ : $\N$) : Sum(n$_1$, n$_2$, n$_3$) $\land$ Sum(n$_2$, n$_1$, n$_3$)  by induction :
case Zero ->
|
|---------------------------------------------------------------------------
| 1: prove $\forall$ (n$_2$ : $\N$) : $\exists$ (n$_3$ : $\N$) : Sum(Zero, n$_2$, n$_3$) $\land$ Sum(n$_2$, Zero, n$_3$)

case S(m$_1$) ->
| ind-hyp: $\forall$ (n$_2$ : $\N$) : $\exists$ (n$_3$ : $\N$) : Sum(m$_1$, n$_2$, n$_3$) $\land$ Sum(n$_2$, m$_1$, n$_3$)
|----------------------------------------------------------------------------------------------
| 1: ^\slf^| for any (N$_2$ : $\N$)
|    |------------------------------------------------------------------------------------------
|    | $\exists$ (n$_3$ : $\N$) : Sum(S(m$_1$), N$_2$, n$_3$) $\land$ Sum(N$_2$, S(m$_1$), n$_3$)  by case analysis on N$_2$ :
|    | case Zero ->
|    | |
|    | |--------------------------------------------------------------------
|    | | 2: prove  $\exists$ (n$_3$ : $\N$) : Sum(S(m$_1$), Zero, n$_3$) $\land$ Sum(Zero, S(m$_1$), n$_3$)
|    |
|    | case S(m$_2$) ->
|    | |
|    | |----------------------------------------------------------------------------------------
|    | | 3: $\exists$ (n$_3$ : $\N$) : Sum(m$_1$, S(m$_2$), n$_3$) $\land$ Sum(S(m$_2$), m$_1$, n$_3$) by rule $\forall$-elim on ind-hyp
|    | | 4: ^\hspace{-2.9pt}^| p: Sum(m$_1$, S(m$_2$), N$_3$) $\land$ Sum(S(m$_2$), m$_1$, N$_3$) for some (N$_3$ : $\N$)
|    | |    |------------------------------------------------------------------------------------
|    | |    | ^\hc{5: Sum(m$_1$, S(m$_2$), N$_3$) by rule $\land$-elim on p}^
|    | |    | ^\hc{6: Sum(S(m$_2$), m$_1$, N$_3$) by rule $\land$-elim on p}^
|    | |    | ^\hc{7: Sum(S(m$_1$), S(m$_2$), S(N$_3$)) by rule sum-s on 5}^
|    | |    | 8: $\forall$ (n$_1$,n$_2$,n$_3$ : $\N$): Sum(n$_1$, n$_2$, n$_3$)$\implies$Sum(n$_1$, S(n$_2$), S(n$_3$)) by theorem sum-s-rhs
|    | |    | 9: Sum(S(m$_2$), m$_1$, N$_3$) $\implies$ Sum(S(m$_2$), S(m$_1$), S(N$_3$)) by rule $\forall$-elim on 8
|    | |    | ^\hc{10: Sum(S(m$_2$), S(m$_1$), S(N$_3$))}^ by rule $\implies$^\hspace{-3pt}^-elim on 9, 6
|    | |    | ^\hc{11: Sum(S(m$_1$), S(m$_2$), S(N$_3$)) $\land$ Sum(S(m$_2$), S(m$_1$), S(N$_3$)) by rule $\land$-intro on 7, 10}^
|    | |    | 12: $\exists$ (n$_3$: $\N$) : Sum(S(m$_1$), S(m$_2$), n$_3$) $\land$ Sum(S(m$_2$), S(m$_1$), n$_3$) by rule $\exists$-intro on 11
|    | | 13: $\exists$ (n$_3$ : $\N$) : Sum(S(m$_1$), S(m$_2$), n$_3$) $\land$ Sum(S(m$_2$), S(m$_1$), n$_3$) by rule $\exists$-elim on 3, 4
| 14: $\forall$ (n$_2$ : $\N$) : $\exists$ (n$_3$ : $\N$) : Sum(S(m$_1$), n$_2$, n$_3$) $\land$ Sum(n$_2$, S(m$_1$), n$_3$) by rule $\forall$-intro on 1
\end{lstlisting}
\end{mdframed}
\caption{A semi-manual formal proof of the theorem \textit{sum-total-comm}}
\label{fig:formal}
\end{figure}
\noindent
Despite the significant verbosity of the notation\footnote{Focusing on the highlighted parts of the most nested sub-proof makes it more readable. It also roughly corresponds to the level of verbosity we would want in future versions of the tool.} (notably when dealing with $\forall$ and $\exists$ eliminations), the proof shows the absence of the uninteresting parts and puts all of the focus on the part considered interesting and requiring creativity. But could it be constructed in a way where the user only has to explicitly write the highlighted interesting parts?

\subsection{Formal Proof in Mixed-Initiative Mode}
\label{sec:interactive}

In the standard checking mode of interaction, the tool handles simple parts of the proof when the user writes the rest of the proof manually. The aim of the alternative mixed-initiative mode of interaction that we present in this paper is to reverse the way of working -- ask the tool to prove the whole theorem on its own and only seek human input when it is needed.

In the mixed-initiative mode, the proof assistant can not only complete the simple cases, but
it is also able to generate the structure of the proof. Depending on the specific configuration of the proof-search strategy (discussed in Section~\ref{sec:Search}), the tool deals with some parts automatically and prompts the user when it gets stuck.

For the example proof discussed in this section, our proof assistant finds a structure that matches the one written manually and it handles those parts that were left for automatic proof-search in the original proof above. The interaction with our prototype is started by entering the following command in the context of a proof:

\vspace{0.5em}
\texttt{prove $\forall$ (n$_1$ : $\N$) (n$_2$ : $\N$) : $\E$ (n$_3$ : $\N$) : Sum(n$_1$, n$_2$, n$_3$) $\land$ Sum(n$_2$, n$_1$, n$_3$)}
\vspace{0.5em}

\noindent
the tool starts searching for a way to justify the given goal.
Depending on the specific configuration (in the case of our prototype -- depending on the hardcoded strategy) it might be able to deal with parts of the proof on its own and only get stuck when it needs to prove\\
\texttt{$\exists$ (n$_3$ : $\N$) : Sum(S(m$_1$), S(m$_2$), n$_3$) $\land$ Sum(S(m$_2$), S(m$_1$), n$_3$)}. At that point, it prompts the user with a request for help. In our prototype, the output looks like this:
\footnote{In our current version the interaction is much more verbose. The output of the tool has been shortened for clarity.}

\vspace{0.75em}
\begin{mdframed}[backgroundcolor=light-gray, roundcorner=10pt,leftmargin=0, rightmargin=0, innerleftmargin=0, innertopmargin=0,innerbottommargin=0, outerlinewidth=0, linecolor=light-gray]
\begin{lstlisting}[mathescape=true,columns=flexible,breaklines=true,breakatwhitespace=true,escapechar=^]
I am solving a theorem `sum-total-comm'

1)  goal: `$\forall$ (n$_1$ : $\N$) : $\forall$ (n$_2$ : $\N$) : $\exists$ (n$_3$ : $\N$) : Sum(n$_1$, n$_2$, n$_3$) $\land$ Sum(n$_2$, n$_1$, n$_3$)'
    strategy: induction

2)  case n$_1$ := S(m$_1$)
    goal: `$\forall$ (n$_2$ : $\N$) : $\exists$ (n$_3$ : $\N$) : Sum(S(m$_1$), n$_2$, n$_3$) $\land$ Sum(n$_2$, S(m$_1$), n$_3$)'
    ind. hypothesis: `ind-hyp$_1$ : $\forall$ (n$_2$ : $\N$) : $\exists$ (n$_3$ : $\N$) : Sum(m$_1$, n$_2$, n$_3$) $\land$ Sum(n$_2$, m$_1$, n$_3$)'

  3)  goal: `$\forall$ (n$_2$ : $\N$) : $\exists$ (n$_3$ : $\N$) : Sum(S(m$_1$), n$_2$, n$_3$) $\land$ Sum(n$_2$, S(m$_1$), n$_3$)'
      strategy: induction

  4)  case n$_2$ := S(m$_2$)
      goal: `$\exists$ (n$_3$ : $\N$) : Sum(S(m$_1$), S(m$_2$), n$_3$) $\land$ Sum(S(m$_2$), S(m$_1$), n$_3$)'
      ind. hypothesis: `ind-hyp$_2$ : $\exists$ (n$_3$ : $\N$) : Sum(S(m$_1$), m$_2$, n$_3$) $\land$ Sum(m$_2$, S(m$_1$), n$_3$)'

My goal is: `$\exists$ (n$_3$ : $\N$) : Sum(S(m$_1$), S(m$_2$), n$_3$) $\land$ Sum(S(m$_2$), S(m$_1$), n$_3$)'.
Type a command or a proof:
\end{lstlisting}
\end{mdframed}
\vspace{0.75em}

\noindent
At this point, the user can write the part of the proof from above -- between \textit{3} and \textit{13} (inclusive). Here's that part again: (Mind the only change -- a different name for the induction hypothesis.)

\vspace{0.75em}
\begin{mdframed}[backgroundcolor=light-gray, roundcorner=10pt,leftmargin=0, rightmargin=0, innerleftmargin=0, innertopmargin=0,innerbottommargin=0, outerlinewidth=0, linecolor=light-gray]
\begin{lstlisting}[mathescape=true,columns=flexible,breaklines=true,breakatwhitespace=true,escapechar=^]
3: $\exists$ (n$_3$ : $\N$) : Sum(m$_1$, S(m$_2$), n$_3$) $\land$ Sum(S(m$_2$), m$_1$, n$_3$) by rule $\forall$-elim on ^\hc{ind-hyp$_1$}^
4: ^\hspace{-1pt}^| p: Sum(m$_1$, S(m$_2$), N$_3$) $\land$ Sum(S(m$_2$), m$_1$, N$_3$) for some (N$_3$ : $\N$)
   |-----------------------------------------------------------------------------------------
   | 5: Sum(m$_1$, S(m$_2$), N$_3$) by rule $\land$-elim on p
   | 6: Sum(S(m$_2$), m$_1$, N$_3$) by rule $\land$-elim on p
   | 7: Sum(S(m$_1$), S(m$_2$), S(N$_3$)) by rule sum-s on 5
   | 8: $\forall$ (n$_1$, n$_2$, n$_3$ : $\N$) : Sum(n$_1$, n$_2$, n$_3$) $\implies$ Sum(n$_1$, S(n$_2$), S(n$_3$)) by theorem sum-s-rhs
   | 9: Sum(S(m$_2$), m$_1$, N$_3$) $\implies$ Sum(S(m$_2$), S(m$_1$), S(N$_3$)) by rule $\forall$-elim on 8
   | 10: Sum(S(m$_2$), S(m$_1$), S(N$_3$)) by rule $\implies$^\hspace{-3pt}^-elim on 9, 6
   | 11: Sum(S(m$_1$), S(m$_2$), S(N$_3$)) $\land$ Sum(S(m$_2$), S(m$_1$), S(N$_3$)) by rule $\land$-intro on 7, 10
   | 12: $\exists$ (n$_3$: $\N$) : Sum(S(m$_1$), S(m$_2$), n$_3$) $\land$ Sum(S(m$_2$), S(m$_1$), n$_3$) by rule $\exists$-intro on 11
13: $\exists$ (n$_3$ : $\N$) : Sum(S(m$_1$), S(m$_2$), n$_3$) $\land$ Sum(S(m$_2$), S(m$_1$), n$_3$) by rule $\exists$-elim on 3, 4
\end{lstlisting}
\end{mdframed}
\vspace{0.75em}

\noindent
At this point, the tool accepts the proof and the search for the proof of the theorem succeeds.
The proof checked by the proof assistant, in this case, is the same as in the checking mode of
interaction, but the user had to write a much smaller proportion of the proof code explicitly.

In our current prototype, the tool merely reports success but does not remember the guidance from the user or the discovered proof. However, it could easily output the structure of the proof, either in full or with gaps to be completed automatically. This way, we could construct the proof initially in the mixed-initiative mode, obtaining an artifact that can be automatically and reproducibly checked in the checking mode of interaction.

\section{Search Strategy}
\label{sec:Search}

The current search strategy is intentionally simple and as non-creative as possible. This is to demonstrate that even a simple search configuration, combined with the mixed-initiative mode of interaction, is sufficient for constructing proofs.
The current prototype does not support configurable search strategies so the prototype has the ``simple configuration'' simply hard-corded, but we expect to explore more advanced and creative
strategies in the future. In our prototype, the search works in the following way.

\begin{itemize}[leftmargin=15pt]
\item When the current goal is a literal:
  \begin{itemize}[leftmargin=15pt]
    \item we try to find an exact match (using unification) in the environment.
    \item we try to use a rule (schema) that would justify the corresponding literal. If a rule's conclusion matches (unifies), we attempt to satisfy all the goals introduced by the rule's premises.
    \item we try to use a theorem (or a lemma).
  \end{itemize}

\item When the current goal is a formula with a conjunction respectively disjunction at the top, we split that goal and try to prove both respectively at least one of its parts.

\item When the current goal is an implication, we put the premise of it in the local scope and within that augmented scope we try to prove the conclusion of the implication.

\item When the current goal is an existentially quantified formula, we instantiate it with a fresh unification variable and try to prove the body of the formula.

\item When the current goal is a universally quantified formula, we use induction to prove it. We do case split according to the type of the quantified variable and supply the induction hypothesis accordingly.
\end{itemize}

It is worth pointing out that our current search strategy intentionally does not fully support backtracking. The theorem that we use for demonstration is so simple that had we implemented the search for all supported logical connectives and made the backtracking work it would be able to solve that theorem completely on its own. That would take away the opportunity to showcase the intended behavior on a simple example. With our extremely simple search strategy, we can illustrate the situation when the automated reasoning fails to find the proof on its own and needs help from the human user, that is the mode of interaction that we want to support for proofs about more realistic programming langauge models in the future.

\section{Ideas to Explore}
The prototype proof assistant outlined above raises a number of research questions that we intend
to explore in the future. One set of topics is motivated by our interest in educational use of the
proof assistant, where it needs to support not just construction of proofs, but also their better
understanding. Another set of topics is concerned with the best way of implementing and supporting
the mixed-initiative mode of interaction.

\paragraph{Educational use of the assistant.}
For use as an educational tool, it is important that the system should not become a black box like
a sophisticated ``auto'' tactic. It needs to allow writing the proofs by hand and merely checking
them. The search strategy needs to be simple enough to be understandable. To support learning,
the system may also offer multiple degrees of automation, so that it can be used as a checker
at the start, gradually becoming more automatic.



\paragraph{When to ask for human insight.}
Our experiments suggest that our proof assistant can easily ``take the wrong turn''
and get stuck leading to prompting the user in a situation in which there is no going forward.
In such cases, the user would be forced to backtract the search, defying the point
of having an automated system. An interesting problem to investigate is thus when to ask the
human user for insight -- the tool would ideally do so as soon as it faces a decision that
would lead it to a state that cannot be solved automatically, but detecting such a state is not
trivial.

A partial solution would be to propose steps and ask the user for confirmation
or offer a choice of multiple directions to explore. This could, however, lead to the user micro-managing the tool and again, defy the point of having the tool.
Another option would be to rely on configurable tactics. The strategies would be able to specify whether it is necessary or preferred to prompt the user before they make a certain action. Finally, a visual interface might also help with this issue.

It might be possible to avoid the issue of choosing between prompting way too late or way too often by changing the interaction a little bit. Instead of the tool asking for help when it gets stuck and the user being presented with a prompt reconstructing the line of reasoning that led to the tool getting stuck, the user would start from the goal and let the tool retrace the path leading to getting stuck. So, instead of the tool asking for permission to do a step too often or asking for forgiveness too late, it allows the user to look at the attempt in a similar way to how a teacher would look at an incorrect solution from a student.
We intend to explore those and similar ideas and try and combine them into a design that would offer the best fitting strategy for each specific situation.

\paragraph{Configurable compositional tactics.}
One way of thinking about automation is as a composition of simple search tactics.
This may be useful as an implementation technique, but also as a way to support the educational
use outlined above. We imagine the system may be configured by specifying a structure consisting
of pre-defined composable more primitive tactics. Basic automation for a novice user may include
just a basic search. A more sophisticated configuration may invoke multiple automatic or interactive
tactics.

It seems that some of the directions with respect to interactivity mentioned above would go well
with a Prolog-like way of defining the proof search.  The author of the tactics library
(the user or a teacher) would be able to specify one or more ways to handle a goal having a certain shape or a certain quality.
The search procedure would then proceed similarly to Prolog languages -- in a depth-first way with backtracking.

\paragraph{Notation and user interaction.}
Currently, the tool is implemented as a terminal utility that prints information and accepts
textual input, but we want to investigate multiple ways of displaying information and
interacting with the tool.
There are two basic modes of interaction. First, the assistant can use an interactive
command-based interface where some state is displayed to the user and the user enters commands
to alter it. Second, the assistant can be built around a representation of proof that
it gradually evolves and that the user can manually change. (The first style resembles
systems like Coq, whereas the latter systems like Agda.) Both cases can support
the mode of interaction we propose where the automatic search can ask for user input.
It may also be interesting to embed the interaction in a structured editor (e.g.
\cite{moon-2022-tylr,omar-2017-hazelnut,beckmann-2023-all}),
especially in the second case (interaction built around a representation of a proof).

\section{Conclusions}
We believe that more powerful tools supporting the work of programming language researchers can
be created by a closer integration between semantics engineering tools, which make it possible to
define, explore and debug formal models of programming languages, and proof assistants, which
can be used to formalize and check properties of such formal models.

This paper is one small step towards our vision. We believe that proof assistants, used by
programming language researchers, can be made more effective by using mixed-initiative mode
of interaction, where the assistant performs simple automatic search and the human provides
crucial insight at the right time. We illustrated this way of working using a simple proof,
checked using our prototype proof assistant. Although our prototype is basic (and could easily
be made more powerful and less verbose), it demonstrates the mode of interaction that we envision.

In the future, we believe the mixed-initiative mode of interaction could leverage other insights
about the structure of programming language models, including those about the typical strucutre
of their proofs, but also insights gained from interactive exploration of programming language
models, such as those done in semantics engineering tools today.

\begin{acks}
The research has been supported by the \grantsponsor{PRIMUS}{Charles University}{} PRIMUS grant
\grantnum{PRIMUS}{PRIMUS/24/SCI/021}. We thank to anonymous reviewers at HATRA'24 for valuable
feedback and references.
\end{acks}

\bibliography{bibfile}



\end{document}